\documentclass[aps,pre,twocolumn,showpacs,superscriptaddress,groupedaddress]{revtex4}  
\usepackage{graphicx}  
\usepackage{dcolumn}   
\usepackage{bm}        
\usepackage{amssymb}   
\usepackage{amsmath}
\usepackage{xcolor}

\definecolor{darkred}{rgb}{.8,0,0}

\definecolor{darkblue}{rgb}{0,0,.7}

\usepackage{amsthm}
\usepackage{enumitem}
\begin{document}


\title{Comment on ``R\'{e}nyi entropy yields artificial biases not in the data and incorrect updating due to the
finite-size data''}

\author{Petr Jizba}
\email{p.jizba@fjfi.cvut.cz}
\affiliation{FNSPE,
Czech Technical University in Prague, B\v{r}ehov\'{a} 7, 115 19, Prague, Czech Republic}

\author{Jan Korbel}
\email{jan.korbel@meduniwien.ac.at}
\affiliation{Section for Science of Complex Systems, Medical University of Vienna, Spitalgasse 23, 1090 Vienna, Austria}
\affiliation{Complexity Science Hub Vienna, Josefst\"{a}dter Strasse 39, 1080 Vienna, Austria}
\affiliation{FNSPE,
Czech Technical University in Prague, B\v{r}ehov\'{a} 7, 115 19, Prague, Czech Republic}

\begin{abstract}
In their recent paper [Phys. Rev. E 99 (2019) 032134], T.~Oikinomou and B.~Bagci have argued that
R\'{e}nyi entropy is ill-suited for inference purposes because it is not consistent with the Shore--Johnson axioms
of statistical estimation theory.
In this Comment we seek to clarify the latter statement by showing that there are several issues in Oikinomou--Bagci reasonings which lead to erroneous conclusions. When all these issues are properly accounted for, no violation of Shore--Johnson axioms is found. 
%
%
\end{abstract}
\pacs{05.20.-y, 02.50.Tt, 89.70.Cf}
\keywords{Shore--Johnson axioms, R\'{e}nyi Entropy, estimation theory}

\vspace{-1mm}
\maketitle
%
%
\noindent
{\em {Introduction}.}~---~
Maximum entropy (MaxEnt) principle  belongs among the most prominent concepts of contemporary statistical physics,
information theory and statistical estimation. Its inception dates back to two seminal papers of E.T.~Jaynes~\cite{jaynes1,jaynes2} who first employed the Shannon information measure, or Shannon entropy (SE), in the framework of equilibrium statistical physics.

Over the years, Jaynes' heuristic MaxEnt prescription has become a powerful instrument, e.g., in non-equilibrium statistical physics, astronomy, geophysics, biology, medical diagnosis or economics~\cite{thurner-book,kapur}.
The rationale behind this success is typically twofold: first, maximizing entropy minimizes the
amount of prior information built into the distribution (i.e. MaxEnt distribution is maximally noncommittal
with regard to missing information); second, many physical systems tend to move towards (or concentrate extremely close to) MaxEnt configurations over time~\cite{Tsallis-book,jaynes1,beck,thurner-book}.

With the advent of generalized entropies~\cite{jizba3,tsallis2,havrda,sharma,Kaniadakis,hanel,korbel}, a natural question has arisen as to whether the MaxEnt principle can be extended also to non-Shannonian entropies. This clearly cannot be decided within Jaynes' heuristic framework --- a sound mathematical qualification is needed.
Since the MaxEnt principle is in its essence an inference method estimating the probability distributions
from limited information, a pertinent mathematical basis should stem from theory of statistical
estimation. Shore and Johnson (SJ)~\cite{shore,shoreII} introduced a system of axioms, which ensure that the MaxEnt estimation procedure is consistent with desired properties of inference methods.
These axioms are as follows~\cite{shore,shoreII}:
\begin{enumerate}
\item \emph{uniqueness}: the system should be unique;
\item \emph{permutation invariance}: the permutation of states should not matter;
\item \emph{subset independence}: It should not matter whether one treats disjoint subsets of system states in terms of separate conditional distributions or in terms of
the full distribution;
\item \emph{system independence}: It should not matter whether one accounts for independent constraints related to independent systems separately in terms of marginal distributions or in terms of full-system.
\end{enumerate}
There is often also a fifth axiom, which was not included in the original systems of SJ axioms~\cite{shore} but
appeared in later editions~\cite{uffink}:
\begin{enumerate}
\setcounter{enumi}{4}
\item \emph{maximality}: In absence of any prior information, the uniform distribution should be the solution.
\end{enumerate}
One can analogously define the set of axioms for the continuous systems with several adjustments~\cite{shore,uffink}. {\em First}, for the continuous variables it is necessary to use the Minimum Relative Entropy
(MinRel) principle,
where the maximization of the entropy is replaced by minimization of the relative entropy subject to some given prior distribution. {\em Second}, the second axiom changes
to \emph{coordinate invariance} axiom, which states that the change of coordinate system should not matter. {\em Third}, the maximality axiom is replaced by \emph{no-information} axiom in the way that in the
absence of any information the prior distribution remains unchanged.  The MaxEnt principle then represents a special case of the MinRel principle for discrete variables and uniform prior distribution.

In recent years, there has been much debate as to whether generalized entropies can fulfill SJ axioms, and if yes, how the permissible classes are classified  (see, e.g.,~\cite{presse,tsallis,presse2} and citations therein).
In their latest paper~\cite{bagci},  Oikinomou and Bagci (OB) focused on the particular case of R\'{e}nyi's
entropy (RE) and argued that RE is not consistent with some of SJ axioms
and hence it is ill-suited for inference purposes. This finding is, however, at odds with recently found one-parameter class of (entropic) functionals --- so-called Uffink's class, which is consistent with SJ axioms~\cite{jizba19},
and which contains RE as a particular member.
In addition, if the OB statement was true, then in some important cases, such as in the R\'{e}nyi entropy-based signal processing and pattern recognition, there would be important new corrections or inconsistencies to some existing
analyzes. OB result would be also detrimental in quantum information theory~\cite{puchala,Zyczkowski} (e.g., RE for $q=2$ is related to \emph{purity}). Here we show that the RE as they stand is {\em certainly} compatible with SJ axioms.
Rather than appealing to Ref.~\cite{jizba19} for a full-fledged proof of  Uffink's class, we will employ more straightforward approach.
In particular, in this Comment we directly point out several issues in OB reasonings. We carefully go through OB arguments and correct respective problematic points. When all issues are properly accounted for, no violation of SJ is
found. For the sake of simplicity, we focus on the discrete version of SJ axioms and change
to continuous variables only when necessary.

In the following we will denote the RE of order $q$ as
\begin{equation}
H_q(P) \ = \ \frac{1}{1-q} \ln \left( \sum_i p_i^q \right), \;\;\;\; q > 0\, ,
\end{equation}
and the ensuing relative RE (or R\'{e}nyi's divergence of order $q$)~\footnote{Notice that the relative RE exists in two versions (both proposed by R\'{e}nyi himself~\cite{Renyi}).
In our reasonings here we stick
to 1st R\'{e}nyi's version which is also employed in the OB paper.}, as
\begin{equation}
H_q(P||Q) \ = \ \frac{1}{q-1} \ln \left\{ \int  \left[\frac{p({\boldsymbol{x}})}{q({\boldsymbol{x}})}\right]^q \!\!q({\boldsymbol{x}}) \ \! \mathrm{d} {\boldsymbol{x}}  \right\} .
\label{RREa}
\end{equation}
%


\noindent
{\em {Critical revision of the OB paper}.}~---~
Let us now go step-by-step through the key arguments presented by OB in~\cite{bagci}. Our discussion will be organized
in the descendent order according to respective SJ axiomatic points:

\paragraph*{1. Uniqueness axiom ---} OB conclude that the first axiom is fulfilled only for $q \in (0,1)$, since only for such values is RE concave. The latter is certainly true, however concavity
is only {\em sufficient} not {\em necessary} condition for uniqueness. A key observation in this context is
that RE is a strictly Schur-concave function for arbitrary $q>0$ \footnote{In general, when a given function $H$ is strictly Schur-concave this means that for two probability vectors satisfying the majorization relation
$p\prec q$, (i.e., $\sum_{i=1}^k p_{(i)} \leq \sum_{i=1}^k q_{(i)}$ for $k \in \{1,\dots,n-1\}$, where $p_{(i)}$ and $q_{(i)}$ are the ordered distributions components in the descending order) one has the inequality $H(q) \geq H(p)$ where $H(p) = H(q)$ only if $p=q$. It should be stressed that a Schur-concave function need not be concave  and concave function need not be Schur-concave. However, a symmetric concave function is always Schur-concave. Both concepts of concavity and Schur-concavity coincide for functions of one variable.} which is also sufficient condition cf.~\cite{puchala,Zyczkowski}. Let us assume that there exist two distinct distributions $P_1$ and $P_2$ (which are not permutation of each other) that maximize $H_q$ under given constraints. Let us take a convex combination $P_\alpha = \alpha P_1 + (1-\alpha) P_2$. Indeed, $P_\alpha$ belongs to the probability simplex and also fulfills the constraints. Since $H_q$ is strictly Schur-concave, it fulfills the following inequality (see also \cite{roberts})
\begin{eqnarray}
H_q(P_\alpha) > \alpha H_q(P_1) + \geq (1-\alpha) H_q(P_2)\nonumber\\
=  H_q(P_1) =  H_q(P_2)\, .
\end{eqnarray}
Thus, the result must be unique otherwise we get contradiction with maximality assumption~\cite{uffink}. This fact will also be important in connection with {\em subset independence} axiom.

\paragraph*{2. Invariance axiom ---} For discrete case, the permutation invariance axiom means that the entropy should be symmetric function of probabilities, which is indeed the case for RE since RE is Schur-concave.
Let us recall that every concave and symmetric function is Schur-concave. The opposite implication is not true, but all Schur-concave functions (including RE) are symmetric (under permutation of the arguments)~\cite{Wayne}.

For continuous variables, one should use (\ref{RREa}). The latter is manifestly invariant under the change of coordinate system ${\boldsymbol{x}} \mapsto {\boldsymbol{y}}$.
Indeed, if ${\boldsymbol{y}} = {\boldsymbol{\varphi}}({\boldsymbol{x}})$ and ${\boldsymbol{\varphi}}$ is a bijective, differentiable function then the well known transformation
rule for probability density functions~\cite{feller}  states that
$p_{Y}({\boldsymbol{y}}) = p_{X}({\boldsymbol{x}}) |\det(\partial {{\boldsymbol{\varphi}}}^{-1}/\partial {\boldsymbol{y}})|$. By setting $p({\boldsymbol{x}})\equiv p_{X}({\boldsymbol{x}})$
and $q({\boldsymbol{x}})\equiv q_{X}({\boldsymbol{x}})$, and plugging this to (\ref{RREa}) wee see that the latter is invariant under the change ${\boldsymbol{x}} \mapsto {\boldsymbol{y}}$.
One could even be more general and employ Radon--Nikodym theorem~\cite{RDa}. With this
the R\'{e}nyi's divergence of order $q$ from $P$ to $Q$ can be rewritten as
\begin{equation}
H_q(P||Q) \ = \ \frac{1}{q-1} \ln \left[ \int  \left(\frac{dP}{dQ}\right)^{q-1} \ \! \mathrm{d} P \right] .
\label{RREb}
\end{equation}
where $dP/dQ$ is the Radon--Nikodym derivative. In this formulation is $H_q(P||Q) $ manifestly coordinate-system independent.

\paragraph*{3. Subset independence axiom ---} Here OB argue that the RE does not fulfill the subset independence axiom. To support their claim they use the Livesey--Skilling criterion~\cite{livesey}:
Any inference (entropic) functional is consistent with subset
independence axiom if for $j \neq k \neq l$ the following identity holds
\begin{equation}\label{eq:crit}
\frac{\partial}{\partial p_l} \left(\frac{\partial}{\partial p_k} - \frac{\partial}{\partial p_j}\right) \left[H - \alpha \sum_i p_i - \beta \sum_i E_i p_i\right]  =  0\, .
\end{equation}
By using (\ref{eq:crit}), OB show that RE does not fulfill this criterion and therefore does not conform with the subset independence axiom.
It is, however, not difficult to see that (\ref{eq:crit}) does not cover all possible configurations and as it stands it is too restrictive.
In fact, in~\cite{shore} has been shown that the entropic functional satisfying first three SJ axioms
must be of the form
\begin{equation}\label{eq:sum}
S(P) \ = \ f\left(\sum_i g(p_i)\right),
\end{equation}
where $f$ is an arbitrary increasing function and $g$ is an increasing concave function. Above entropic functionals are called \emph{sum-form} entropies. The proof can be found in the Supplemental material of~\cite{jizba19}.
Note that the explicit form of a function $f$ in (\ref{eq:sum}) does not influence the form of the distribution estimated by the MaxEnt principle. This can be easily
seen by comparing two situations:
a) $f(x)= a x$ (with $a > 0$ being a constant), in which case the MaxEnt principle dictates that we
should maximize $\sum_i g(p_i)$ subject to constraints $\sum_i p_i =1$ and $\sum_i p_i E_i = E$. This gives
%
%
\begin{equation}
g'(p_i) \ - \ \alpha \ - \ \beta E_i \ = \ 0\, ,
\end{equation}
and therefore
\begin{eqnarray}
  \alpha &=& \sum_i p_i \ \!g'(p_i) - \beta E\, ,\\[1mm]
  \beta &=& \frac{g'(p_i) - \alpha}{E_i}\, ,\\[1mm]
   p_i &=& (g')^{-1}\left(\alpha \ + \ \beta E_i\right),
\end{eqnarray}
b) $f\neq a x$, where the MaxEnt principle prescribes that we
should maximize $f(\sum_i g(p_i))$ under the same constraints as above. In this case we have
\begin{equation}
C_f g'(p_i) \ - \ \alpha_f \ - \ \beta_f E_i \ = \ 0\, ,
\end{equation}
where  $C_f \equiv f'\left(\sum_i g(p_i)\right)$. This leads to
\begin{eqnarray}
  &&\mbox{\hspace{-5mm}}\alpha_f \ = \ C_f \sum_i p_i g'(p_i) - \beta_f E \ = \ C_f \alpha\, ,\\[1mm]
  &&\mbox{\hspace{-5mm}}\beta_f \ = \ \frac{C_f g'(p_i) - \alpha_f}{E_i} \ = \ C_f \beta\, ,\\[1mm]
   &&\mbox{\hspace{-5mm}}p_i \ = \ (g')^{-1}\left(\frac{\alpha_f \ + \ \beta_f E_i}{C_f}\right)\nonumber \\[1mm]
   &&\mbox{\hspace{-0.5mm}} = \ (g')^{-1}\left(\alpha \ + \ \beta E_i\right),
\end{eqnarray}
so the resulting MaxEnt distribution is indeed independent of $f$. This defines the equivalent classes of entropic functionals with equivalence $f(\sum_i g(p_i)) \sim \sum_i g(p_i)$.

Let us now go back to the criterion (\ref{eq:crit}) and apply it to the class of entropic functionals (\ref{eq:sum}). The difference of derivatives gives
\begin{equation}
\frac{\partial}{\partial p_l}\Big\{ C_f(P) [g'(p_k)-g'(p_j)] \ + \ \beta (E_k - E_j) \Big\},
\end{equation}
and successive derivative with respect to $p_l$ then yields
\begin{eqnarray}
&&\mbox{\hspace{-7mm}} \frac{\partial C_f(P)}{\partial p_l} [g'(p_k)-g'(p_j)]\nonumber \\[1mm]
&& \mbox{\hspace{10mm}}+ \ C_f(P) \frac{\partial}{\partial p_l} \Big[g'(p_k)-g'(p_j)\Big] \ = \ 0\, .
\label{imp}
\end{eqnarray}
The second term on the LHS vanishes for $l \neq j,k$. The first term is zero only when ${\partial C_f(P)}/{\partial p_l} = 0$, which implies that  $f''(x) = 0$, or equivalently $f(x) = ax$.
Consequently, we see that the Livesey--Skilling criterion employed by OB can support only  the \emph{trace-class} entropies, i.e., entropy functionals of the form $\sum_i g(p_i)$.
On the other hand, since the function $f$ does not change the resulting MaxEnt distribution, all sum-form entropies must be consistent with the {\em subset independence} axiom.
The only quantities that are changed are the Lagrange parameters. The transform $\sum_i g(p_i) \mapsto f\left(\sum_i g(p_i)\right)$
can therefore be interpreted as a kind of {\em gauge} invariance in the MaxEnt principle.

Let us finally make two remarks regarding the aforementioned gauge invariance:
\begin{itemize}
\item Since the function $f$ can be arbitrary, it is irrelevant whether the entropic functional is additive or not. Actually, by application of appropriate function, one can impose the desired type of
(generalized) additivity. For example, the RE is additive, while Tsallis entropy~\cite{tsallis2}, $S_q =  \ln_q \exp H_q$ is $q$-additive (here $\ln_q$ is the $q$-deformed
logarithm~\cite{borges}) and RE power~\cite{jizba}, $P_q = \exp H_q$ is multiplicative.
\item From the sum-form (\ref{eq:sum}) of entropy  it is also clear that concavity cannot be necessary condition for uniqueness, since an increasing function of a concave function does not need to be concave.
On the other hand, an increasing function of any Schur-concave function is yields again Schur-concave function~\cite{schur,Wayne}.
\end{itemize}

\paragraph*{4. System independence axiom ---} OB also conclude that RE does not fulfill system independence axiom. Their result relies on the solution of functional equation
\begin{equation}\label{eq:17}
g'''(p_{ij}) \frac{\partial p_{ij}}{\partial u_i} \frac{\partial p_{ij}}{\partial v_j} \ + \ g''(p_{ij}) \frac{\partial^2 p_{ij}}{\partial u_i \partial v_j} \ = \ 0\, ,
\end{equation}
where $p_{ij}$ is the joint distribution of the whole system and $u_i$ and $v_j$ are the marginal distributions of two disjoint subsystems. At this point, OB follow the approach
of Press\'{e} {\em et al.}~\cite{presse} and assume the form of the joint distribution as $p_{ij} = u_i v_j$. From this, they end up with the equation $x g'''(x) + g''(x) = 0$ leading to
$g(x) = -x \log x$ modulo multiplicative and additive constant, which corresponds to Shannon entropy. However, as already discussed in~\cite{uffink,jizba19}, the assumption on the structure of probability distribution goes
well beyond the original idea of consistency axioms, since assume only certain structure of updating information, not the probability distribution itself. In Ref.~\cite{jizba19} it was
pointed out that this requirement can be ensured by assuming a stronger version of the system independence axiom, which can be formulated as follows: whenever two subsystems of a
system are disjoint, we can treat the subsystems in
terms of independent distributions. We shall note that this \emph{strong system independence} axiom is fulfilled for many systems observed in nature, namely for systems which state space scales exponentially. These systems have typically
short-range interactions. In this case, the strong independence axiom allows to bring Eq. \eqref{eq:17} to the following form:
\begin{eqnarray}
\frac{\partial p_{ij}}{\partial u_i} \frac{\partial p_{ij}}{\partial v_j}/\frac{\partial^2 p_{ij}}{\partial u_i \partial v_j}  \ = \ p_{ij}\, .
\end{eqnarray}
On the other hand, this is apparently not the most general form of the relation between joint and marginal distribution. It is beyond the scope of this Comment to investigate the general
form of this relation, but let us just look at the case when the ratio is also linear but with some prefactor.
In this case, we have:
\begin{equation}
\label{eq:com}
\frac{\partial p_{ij}}{\partial u_i} \frac{\partial p_{ij}}{\partial v_j}/\frac{\partial^2 p_{ij}}{\partial u_i \partial v_j} \ = \ a p_{ij}\, ,
\end{equation}
for $a$ close to $1$. This leads to differential equation in the form $a x g'''(x) + g''(x) = 0$ which has the solution $g(x) \propto x^{2-1/a}$. By denoting $q=2-1/a$, i.e., $a= 1/(2-q)$, we end up with entropic functional
equivalent to RE. Eq.~(\ref{eq:com}) corresponds to the composition rule of $q$-exponential distributions
described in~\cite{jizba19}
\begin{eqnarray}\label{eq:18}
p_{ij} \ \propto \ \left(u_i^{q-1} + v_j^{q-1}-1 \right)^{1/(q-1)}\, .
\end{eqnarray}
Reader can easily check that \eqref{eq:18} satisfies Eq. \eqref{eq:17}. As shown in~\cite{uffink} (see also Supplemental Material in Ref.~\cite{jizba19}), RE conforms with the {\em system independence} axiom.

\paragraph*{5. Maximality axiom ---} strict Schur-concavity of RE automatically ensures that in the absence of any prior information, the uniform distribution must be the MaxEnt distribution.
Indeed, $H_q(P) \geq H_q(Q)$ whenever $P \prec Q$ for
any two $n$-dimensional distributions $P$ and $Q$, and hence
\begin{eqnarray}
H_q(1, 0, \ldots, 0)\ \leq \ H_q(P) \ \leq \ H_q(1/n, \ldots, 1/n) \, ,
\label{max5ax.a}
\end{eqnarray}
for any $P$. This is a simple consequence of observation that
\begin{eqnarray}
\{1/n, \ldots, 1/n\} \ \prec \ P \ \prec \ \{1,0,\ldots, 0\}\, .
\label{max5ax.b}
\end{eqnarray}
Note that the lover limit in (\ref{max5ax.a}) is saturated only when $P$ is a permutation of the pure-state distribution $\{1,0,\ldots, 0\}$ and the upper limit is saturated only
for maximally-mixed-state (uniform) distribution $P = \{1/n, \ldots, 1/n\}$. So RE has a strict global maximum at the uniform distribution. This completes the proof. Relations (\ref{max5ax.a})-(\ref{max5ax.b}) also nicely bolster the usual interpretation of entropy --- the larger is the entropy,
the more uniform is the distribution.

\noindent
{\em {The issue of escort constraints}.}~---~
Let us also briefly comment the use of the escort averages $\sum_i \rho_q(p_i) E_i = E_q$ in MaxEnt prescription. It should be first stressed that the whole framework of SJ consistency axioms has been invented
for the case of linear constraints $\sum_i p_i E_i = E$. Its extension to more generalized types of constraints as, for example, escort averages remains an open problem. Therefore, it is not possible to apply the original
SJ criteria to that situation. One possible way how to overcome usage of escort means is to change the probability distribution to the escort distribution, so that the
constraints become linear. In this case, one can also formulate the entropy in terms of escort distributions $p_i = \rho_i^{1/q}/\sum_j \rho_j^{1/q}$, which leads to maximization of
functional equivalent to the Landsberg (or also Homogenous) entropy functional~\cite{lutsko,Landsberg}
\begin{equation}
S_q^H(\rho) \ = \ \frac{\left(\sum_i \rho_i^{1/q}\right)^{-q}-1}{1-q}\, .
\label{landsb}
\end{equation}
When the whole framework is formulated in terms of escort probabilities with linear constraints, we can employ the SJ axioms but formulated in terms of escort distributions. According to \cite{jizba19},
Eq.~(\ref{landsb}) will then belong to the class of Uffink's entropic functionals (i.e., class that is consistent with SJ axioms). This fact alone, however,  does not ensure that the SJ axioms will also be valid
in the original picture. Particularly, the problem might arise in connection with {\em subset independence} axioms, since the de Finneti--Kolmogorov theorem $p_{ij} = u_i v_{j|i}$ does not generally hold for corresponding escort-distribution
counterparts~\cite{jizba17,jizba17b}.

\noindent
{\em {Conclusions}.}~---~
In this Comment we have analyzed the recent claim of T.~Oikinomou and B.~Bagci~\cite{bagci}, that
R\'{e}nyi entropy is ill-suited for inference purposes because it is not consistent with the Shore--Johnson axioms. By carefully examining Oikinomou--Bagci arguments we have, however, noticed that there are several issues in their reasonings that need to be clarified. When the latter are properly accounted for we find that there is no contradiction with SJ desiderata. This conclusion should also be expected on more general ground. Namely, R\'{e}nyi entropy is known to be a bona fide member of the so-called Uffink's class of entropic functionals~\cite{jizba19,uffink}, which is the most general class of inference functionals satisfying SJ axioms~\cite{jizba19}.

Import ingredient in our reasonings was the strict Schur-concavity of R\'{e}nyi entropy.
In fact, the language of majorization and (strict) Schur-concavity is very natural in the context of entropies since
many processes in physics occur in the direction of the majorization arrow (because the passage
of time tends to make things more uniform) and Schur-concave entropies grasp this behavior via their non-decreasing evolution.


\begin{acknowledgements}
{\em {Acknowledgments.}}~---~P.J. and J.K. were supported by the Czech Science Foundation (GA\v{C}R), Grant 19-16066S.
J.K. was also supported by the Austrian Science Foundation (FWF) under project I3073.
\end{acknowledgements}

\end{document}